\documentclass[twocolumn,aps,floatfix,amssymb]{revtex4-1}
\usepackage{amsmath}
\usepackage{graphicx, color}
\usepackage{dsfont}
\usepackage{units}
\usepackage{hyperref}

\renewcommand{\vec}[1]{\mathbf{#1}}
\newcommand{\sref}[1]{section~\ref{#1}}
\newcommand{\eref}[1]{Eq.~(\ref{#1})}
\newcommand{\fref}[1]{Fig.~\ref{#1}}
\newcommand{\Eref}[1]{Equation~(\ref{#1})}
\begin{document}
\title{Switchable topological phonon channels}
\date{\today}
\author{Roman~S\"usstrunk}
\affiliation{Institute for Theoretical Physics, ETH Zurich, 8093 Z\"urich, Switzerland}
\author{Philipp~Zimmermann}
\affiliation{Institute for Theoretical Physics, ETH Zurich, 8093 Z\"urich, Switzerland}
\author{Sebastian~D.~Huber}
\affiliation{Institute for Theoretical Physics, ETH Zurich, 8093 Z\"urich, Switzerland}

\begin{abstract}
	Guiding energy deliberately is one of the central elements in engineering and information processing. It is often achieved by designing specific transport channels in a suitable material. Topological metamaterials offer a way to construct stable and efficient channels of unprecedented versatility. However, due to their stability it can be tricky to terminate them or to temporarily shut them off without changing the material properties massively. While a lot of effort was put into realizing mechanical topological metamaterials, almost no works deal with manipulating their edge channels in sight of applications. Here, we take a step in this direction, by taking advantage of local symmetry breaking potentials to build a switchable topological phonon channel.
\end{abstract}

\maketitle
%
\section{Introduction} 
\label{sec:introduction}
The concept of topological phonon bands emerged recently as a new design principle in mechanical metamaterials \cite{Huber16,Susstrunk16,Kane13,Lubensky15,Paulose15,Paulose15a,Chen15,Meeussen16}. Like its ancestral electronic counterparts \cite{Hasan10,Qi11}, it offers a new approach to shape the transport of energy in materials, which is typically directional and free of backscattering. Furthermore, the channels for the energy transport are extremely stable against a broad range of disorders.

The stability of these channels is granted by the nature of the underlying topology. Consequently, the kinds of disorder the energy transport is immune to depends on the type of topology \cite{Kitaev09,Ryu10}. Some types rely on symmetries \cite{Kane05b,Konig07}, while others exist independently of them \cite{Klitzing80,Thouless82}. There are many realizations of symmetry protected topological effects \cite{Hasan10,Qi11}, but all of them share the property that once the symmetry is broken, quantities building upon the topology loose their stability.

In case that no symmetry is needed, the stability of the edge channels is typically stronger as a larger set of distortions is allowed \cite{Nash15}. For applications this seemingly offers more promising candidates than their symmetry protected counterparts \cite{Susstrunk15,Mousavi15,Xiao15}. However, the superior stability can be become a disadvantage once we want to terminate or temporarily close a channel to influence an energy flow. Here, the symmetry protected systems offer us a more flexible alternative.

By deliberately breaking the symmetry, we have a tool to remove or terminate channels. This is a significant advantage over systems not relying on symmetry, while the price to pay is a reduced stability. Here, we experimentally demonstrate how to use this backdoor to switch on and off a topological phonon channel. While our analysis is based on a specific model, the conclusions carry over to any other system of this type.

The remainder of this paper is structured as follows. We first introduce and discuss the system we are working with (\sref{sec:model}), followed by a theoretical analysis of its eigensolutions (\sref{sec:the_semi_periodic_problem}). In a next step (\sref{sec:observation_of_the_edge_band_gap}), we verify that our setup reproduces the key properties needed for the subsequent experimental implementation of the switch, presented in \sref{sec:switching_a_phonon_channel}.
%
\section{Model} 
\label{sec:model}
As starting point for our studies we use the same setup as in Ref.~\cite{Susstrunk15}: a mechanical version of the quantum spin Hall effect on a discrete periodic square lattice. The band structure of the infinite system supports non-trivial spin Chern numbers \cite{Sheng06}. In a finite system this manifest itself in helical edge states \cite{Kane05a}, allowing for directional phonon channels. The spin Chern numbers and therefore the edge channels are symmetry protected by a local symmetry $S$. Next, we recapitulate some of the systems properties and refer to Ref.~\cite{Susstrunk15} for more details.

\begin{figure}[tbp]
	\centering
		\includegraphics[scale=0.89]{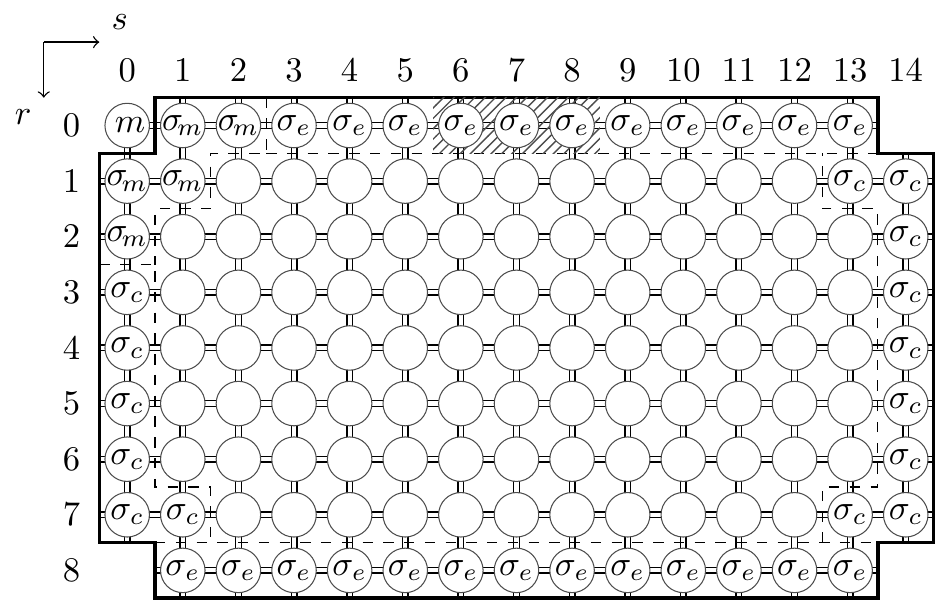}
	\caption{Scheme of the experimental setup, including indication of $r$ and $s$ directions. Each circle represents a site of the system and connecting lines indicate couplings between sites. The corner sites are not part of the dynamical system, but allow us to attach a drive ($m$) to induce a specific motion. The edge row of the system is divided into different sectors $\sigma_{c/e/m}$ needed in the analysis of the experiments. The shaded part marks a region in which the system will be selectively distorted to realize a switch.}
	\label{fig:setup}
\end{figure}

Each lattice site $(r,s)$ consists of two degenerate local modes, $x$ and $y$, which are coupled to the modes of neighboring sites. A unit cell is made up of three sites leading to six degrees of freedom per unit cell. In the experimental realization, the local modes are implemented through one-dimensional pendula which are coupled via springs. The measurements are carried out on a finite system with fixed boundaries, as shown in \fref{fig:setup}. The system size is $9\times 15$ sites, minus the four corner sites which we removed for technical reasons.

The construction is chosen to obtain equations of motion of the form
\begin{equation}\label{eq:eom}
	\begin{aligned}
		\begin{pmatrix}
			\ddot{\vec{x}}_{r,3s} \\ \ddot{\vec{y}}_{r,3s}
		\end{pmatrix}
		&=
		\begin{pmatrix}
			\Omega+{f_r}^x A & 0 \\
			0 & \Omega+{f_r}^y A
		\end{pmatrix}
		\begin{pmatrix}
			{\vec{x}}_{r,3s} \\ {\vec{y}}_{r,3s}
		\end{pmatrix}\\
		&+
		\begin{pmatrix}
			{f_r}^x B^T & 0 \\
			0 & {f_r}^y B^T
		\end{pmatrix}
		\begin{pmatrix}
			{\vec{x}}_{r,3(s- 1)} \\ {\vec{y}}_{r,3(s- 1)}
		\end{pmatrix}\\
		&+
		\begin{pmatrix}
			{f_r}^x B & 0 \\
			0 & {f_r}^y B
		\end{pmatrix}
		\begin{pmatrix}
			{\vec{x}}_{r,3(s+ 1)} \\ {\vec{y}}_{r,3(s+ 1)}
		\end{pmatrix}\\
		&+
		f
		\begin{pmatrix}
			C & \pm D \\
			\mp D & C
		\end{pmatrix}
		\begin{pmatrix}
			{\vec{x}}_{r\pm 1,3s} \\ {\vec{y}}_{r\pm,3s}
		\end{pmatrix},
	\end{aligned}
\end{equation}
where $\vec{z}_{r,3s}=(z_{r,3s},z_{r,3s+1},z_{r,3s+2})^T$, for $z\in\{x,y\}$, correspond to the displacements of the six modes per unit cell. The matrices $\Omega$, $A$, $B$, $C$ and $D$ are real-valued and can be found in the appendix along with the numerical values of the parameters. In the present form, the equations of motion apply to an infinite, periodic or finite system. For the latter we simply set $z_{r,s}=0$ in case site $(r,s)$ is not part of the system. \Eref{eq:eom} is a generalized version of the equations of motion used in \cite{Susstrunk15}, where ${f_r}^{x/y}=f$. This generalization is needed to add an edge potential later on. 

In case that ${f_r}^x=f_r={f_r}^y$, the structure of the coupling elements ensures that
\begin{equation}
	S =
	\left(\hspace{-4pt}\begin{array}{cc}
		0 & \mathds{1}_{3\times 3} \\
		-\mathds{1}_{3\times 3} & 0
	\end{array}\hspace{-4pt}\right),
\end{equation}
is a symmetry of the equations of motion. This symmetry allows us to block-diagonalize the system by transforming the $x$ and $y$ degrees of freedom into left and right polarizations $p_{\pm}$ defined by
\begin{equation}
	{p}_{\pm,r,s} = \frac{1}{\sqrt{2}}({x}_{r,s} \pm i {y}_{r,s})\,.
\end{equation}
In the quantum spin Hall analogue, the polarizations correspond to the two different spins. Through this transformation the equations of motion become block-diagonal and can then be expressed as
\begin{equation}
	\begin{aligned}
		\ddot{\vec{p}}_{\pm,r,3s}
		&=
		(\Omega+f_r A) {\vec{p}}_{\pm,r,3s} \\
		&+
		f_rB {\vec{p}}_{\pm,r,3(s+1)}
		+f_rB^T {\vec{p}}_{\pm,r,3(s-1)} \\
		&+
		f(C\mp i D ){\vec{p}}_{\pm,r+1,3s}+f(C\pm i D ){\vec{p}}_{\pm,r-1,3s}\,.
	\end{aligned}
\end{equation}
The two polarization sectors are related by complex conjugation and are degenerate.

While the system can always be described in a polarization basis, the presence of the symmetry is crucial for the decoupling of the polarization sectors. Without it, the different polarization sectors will couple. The symmetry on the other hand relies on the structure of the equations of motion. More precisely, it depends on the fact that $x$ and $y$ degrees of freedom `see' the same local coupling matrices $\Omega$, ${f_r}^{x/y}A$, ${f_r}^{x/y}B$, $C$ and $D$. If this is no longer given, in particular if ${f_r}^x\neq {f_r}^y$, the two sectors start to mix and the separation breaks down.

%
\section{The (semi-)periodic problem} 
\label{sec:the_semi_periodic_problem}
We start by studying the fully periodic system with ${f_r}^x=f={f_r}^y$. In this case there are no additional edge potentials and the ansatz ${\vec{p}}_{\pm,r,3s} = e^{-i(k_s s+k_r r) +i\omega t}{\vec{p}}_{\pm}$ further simplifies the problem, leading to the dispersion bands
\begin{equation}
	\begin{aligned}
		{\omega_n}^2 
			&= \omega_{0}^{2}+(3+\sqrt{3})f \\
			&\hspace{-4pt}+ \sqrt{8}f\cos\biggl\{\frac{1}{3}\arccos\biggl[-\frac{\cos(3k_r)+\cos(k_s)}{2 \sqrt{2}} \biggr]-\frac{2\pi n}{3}  \biggr\},
	\end{aligned}
\end{equation}
$n\in\{0,1,2\}$. There are three doubly degenerate bands (due to the two polarizations), which are separated by bulk band gaps. By focusing on one polarization only, one can find that the three bands have non-trivial Chern numbers \cite{Thouless82}. Therefore, by cutting the fully periodic system open, edge states must appear in the bulk band gaps due to the bulk-edge correspondence \cite{Fu06,Graf13}.

To create an edge, we consider a system periodic in $s$ direction and semi-infinite in $r$ direction, such that $p_{\pm,r,s}=0$ for $r<0$ \footnote{We could equally well look at an edge in $r$ direction. The choice in our case is due to technical reasons in the experiment.}. This way, there is a single edge along which a wavenumber $k$ can again be introduced due to the periodicity of the system. We then use the ansatz $p_{\pm,r,s}=e^{-is(k\pm 2\pi n_r/3)+i\omega t}p_{\pm,r}$ with $n_{r+1}=n_r-1$ and $n_0\in\{0,1,2\}$. In the symmetry preserved case, this simplifies the equations of motion to
\begin{equation}\label{eq:recursionEdgeSpectrum}
	\begin{aligned}
		p_{\pm,r}\,h_{\pm,r}(\omega,k) 
		& = p_{\pm,r+1}+p_{\pm,r-1}\,, \\
		\phantom{p_{\pm,r}}\,h_{\pm,r}(\omega,k) 
		& = \{(\omega_{0}^{2}-\omega^2)/f+1+\sqrt{3} \\
		&+2f_r/f[1-\cos(k\pm 2\pi n_r/3)]\}\,.
	\end{aligned}
\end{equation}
We note that while $k_s$ and $k$ are along the same direction they differ by a factor of three. The origin of this difference is that the ansatz of the fully periodic problem and the ansatz used in the semi-periodic problem are related by a gauge transformation in the language of the quantum spin Hall effect. The solutions we are looking for have now a periodicity of one unit cell in direction of $s$ at the price of a larger unit cell in the direction of $r$.

\begin{figure*}[tbp]
	\centering
		\includegraphics[scale=1]{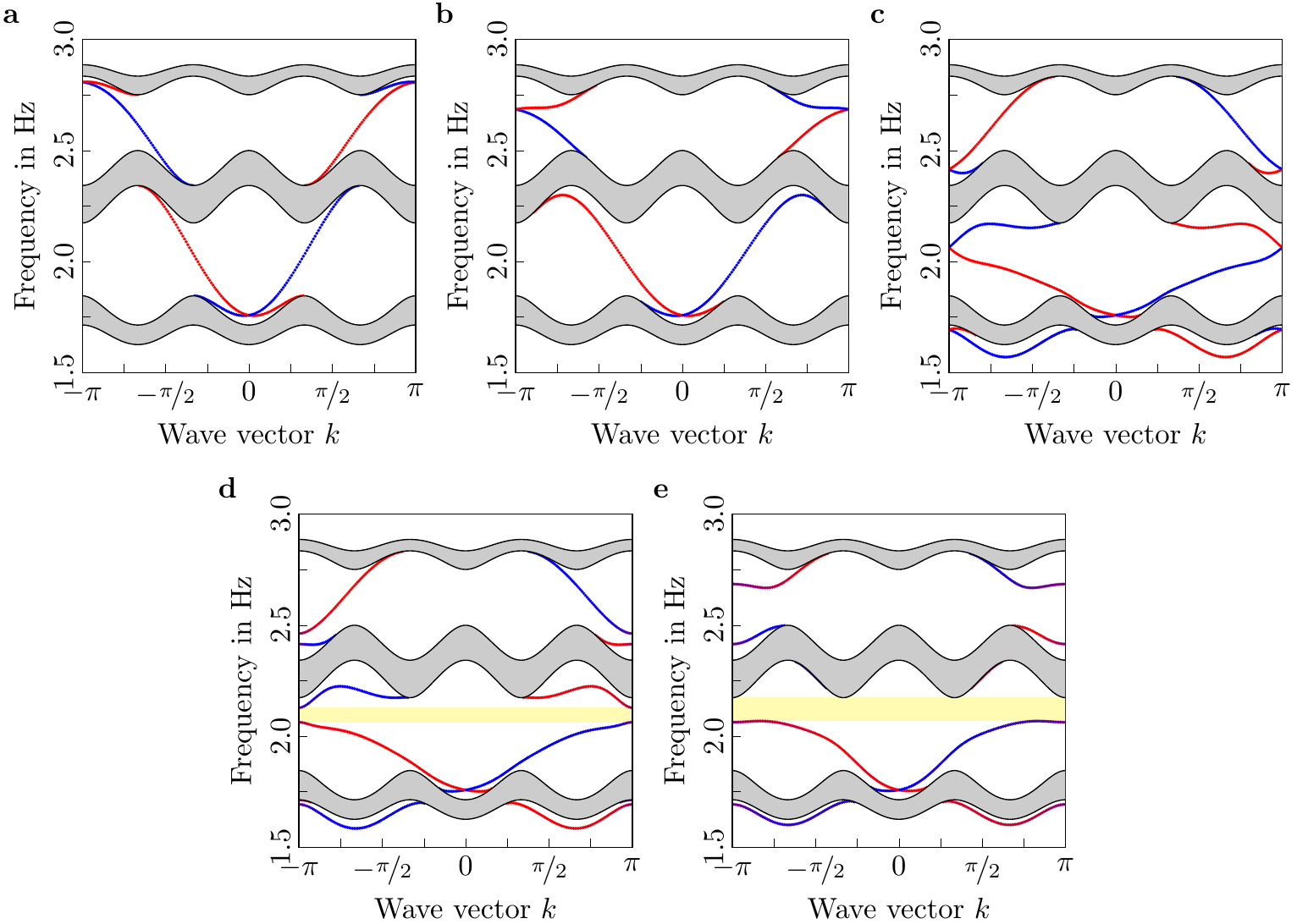}
	\caption{Calculated band dispersions for a semi-periodic system. Gray areas represent bulk bands, while blue (red) lines represent right (left) polarization encoded edge dispersions. (a-c) Dispersions for different symmetry-preserving edge potentials. The chosen potentials are (a) $\alpha=1$, (b) $\alpha=0.81$ and (c) $\alpha=0.24$. (d-e) Dispersions for symmetry-breaking edge potentials, mixing the two polarizations and opening an edge band gap highlighted in yellow. The edge potentials are (d) $\alpha^x=0.40$, $\alpha^y=0.24$ and (e) $\alpha^x=0.81$, $\alpha^y=0.24$.}
	\label{fig:spectra}
\end{figure*}

Choosing an edge potential $f_r/f=\alpha\delta_{r,0}+(1-\delta_{r,0})$, changing only the row at the very edge, provides a handle on the eigensolutions close to the edge. To calculate the spectra we employ the transfer matrix formalism: we reformulate \eref{eq:recursionEdgeSpectrum} as
\begin{equation}\label{eq:transferMatrix}
	\begin{aligned}
		\begin{pmatrix}
			{{p}}_{\pm,r+3} \\ {{p}}_{\pm,r+2}
		\end{pmatrix}
		& =
		T_{\pm}(\omega,k,n_r)
		\begin{pmatrix}
			{{p}}_{\pm,r} \\ {{p}}_{\pm,r-1}
		\end{pmatrix},\qquad r>0\,, \\
		T_{\pm}(\omega,k,n_r)
		&=
		\begin{pmatrix}
			g & 1-h_{\pm,r+1}h_{\pm,r+2} \\ 
			h_{\pm,r}h_{\pm,r+1}-1 & -h_{\pm,r+1}
		\end{pmatrix},\\
		g
		&= h_{\pm,r}[h_{\pm,r+1}h_{\pm,r+2}-1]-h_{\pm,r+2}\,,
	\end{aligned}
\end{equation}
where $T_{\pm}(\omega,k,n_r)=T_{\pm}(\omega,k,n_{r+3})$ denotes the transfer matrix. The requirement $r>0$ is due to the edge potential which makes the row $r=0$ distinct from all the others. The transfer matrix $T_{\pm}(\omega,k,n_1)$ encodes how an eventual solution on the edge (${{p}}_{\pm,0}$ and ${{p}}_{\pm,1}$) evolves into the system. The determinant of the transfer matrix equals $1$, for which reason its two eigenvalues are related by their inverse. Whenever the two eigenvalues are of modulus one, the combination $(\omega,k)$ belongs to a bulk state. However, in case one of the eigenvalues has modulus smaller than one and the corresponding eigenvector is compatible with the edge we are looking at, the combination $(\omega,k)$ represents an edge state \cite{Hatsugai93,Agazzi14}.

This way, the symmetry preserved edge dispersions shown in \fref{fig:spectra} (a-c) are calculated. In contrast to the fully periodic case, the spectrum is no longer gapped. In each bulk band gap there are two edge bands, one for each polarization. The group velocity of each edge band is predominantly either positive or negative. Therefore, the direction of propagation is linked to the polarization of the mode. Through selectively exciting one polarization only, a directional energy transport can be realized.

\begin{figure*}[thb]
	\centering
		\includegraphics[scale=1]{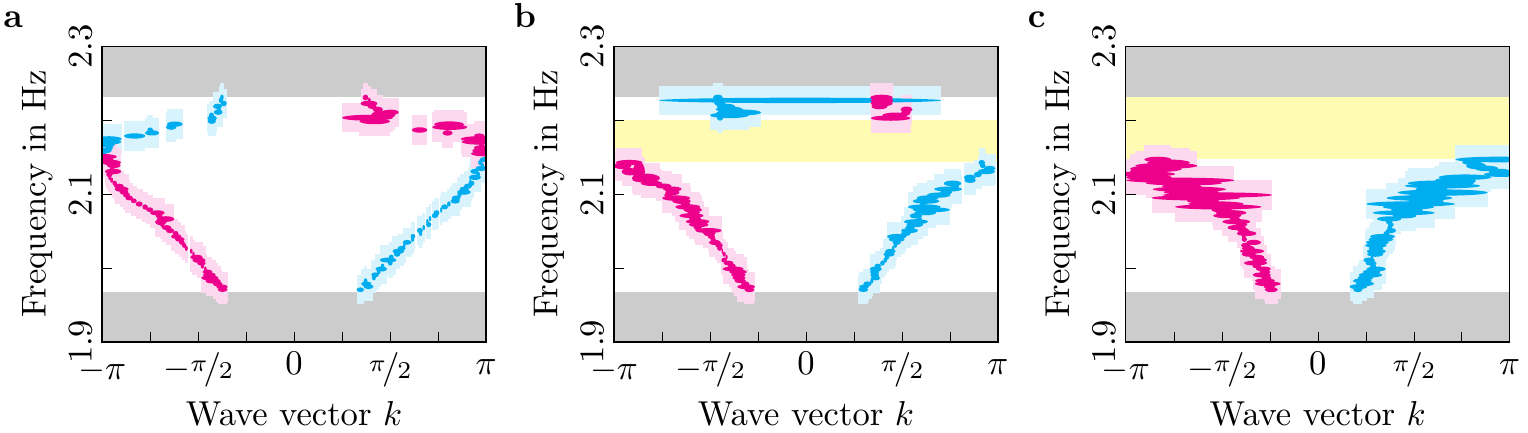}
	\caption{Measured band dispersions. Bulk bands are shaded in gray and total band gaps in yellow. Data points are plotted as ellipses representing their confidence. The lighter shaded areas show the frequency uncertainty due to damping. The blue (red) color encodes the right (left) polarization of the drive during the excitation. (a) Edge dispersion for a symmetry preserving potential as in \fref{fig:spectra} (c). (b-c) Edge dispersions with potential as in \fref{fig:spectra} (d-e), in which polarization of the eigenstates is not well-defined.}
	\label{fig:measuredSpectra}
\end{figure*}

Whenever a symmetry breaking potential, ${f_r}^{x/y}/f={\alpha}^{x/y} \delta_{r,0}+(1-\delta_{r,0})$ with $\alpha^x\neq \alpha^y$,  is applied, the two polarizations are coupled. However, the process of obtaining the edge spectra did not explicitly rely on having the symmetries preserved and can be straightforwardly repeated in absence of the symmetry. The transfer matrix \eref{eq:transferMatrix} then becomes a $4\times 4$ matrix representing the fact that the polarizations cannot be treated independently anymore. We repeated the calculation for the two symmetry broken cases shown in \fref{fig:spectra} (d-e).

The edge band structure can be manipulated through the edge potential. By changing the couplings to lower $\alpha<1$ the crossing points of the edge dispersion can be brought into the bulk band gap. By adding on top a symmetry breaking term, the two polarizations start to mix and we open an edge band gap within the bulk band gap. At frequencies in the total band gap there are no solutions anymore which could support a transport of energy through the system. Combining the symmetry preserved and the symmetry broken systems into one device will eventually lead to the desired switch. Before we turn our attention to that, we look at the experimental implementation of the above band structures.
%
\section{Observation of the edge band gap} 
\label{sec:observation_of_the_edge_band_gap}
The experimental setup is shown in \fref{fig:setup}. The system is finite in all directions in contrast to the above analysis. It provides one big edge, which we further divide into multiple segments. The two short edges $\sigma_c$ along the $r$ direction build a segment and the two long edges $\sigma_e$ along the $s$ direction build one. On top of that, all the sites which have a Manhattan distance of less than three to the drive $m$ are separated into $\sigma_m$. All the unlabeled sites we consider as bulk sites $\sigma_b$.

To access the dispersion relation of the system we harmonically drive it at a given frequency $\omega$. Once a steady state emerged, we measure the time-resolved displacements of all the pendula with a camera. Based on the motion we observe, we can calculate the time averaged energy on every site and with it the mean energy for all sectors. By comparing the energies of different sectors, we can distinguish edge states from bulk states and find gaps in the bulk or edge spectra. Whenever we find an edge state with significant weight in $\sigma_e$, we fit a wavenumber $k$ to the propagation along the edge \cite{Susstrunk15}.

\begin{figure*}[tbp]
	\centering
		\includegraphics[scale=1]{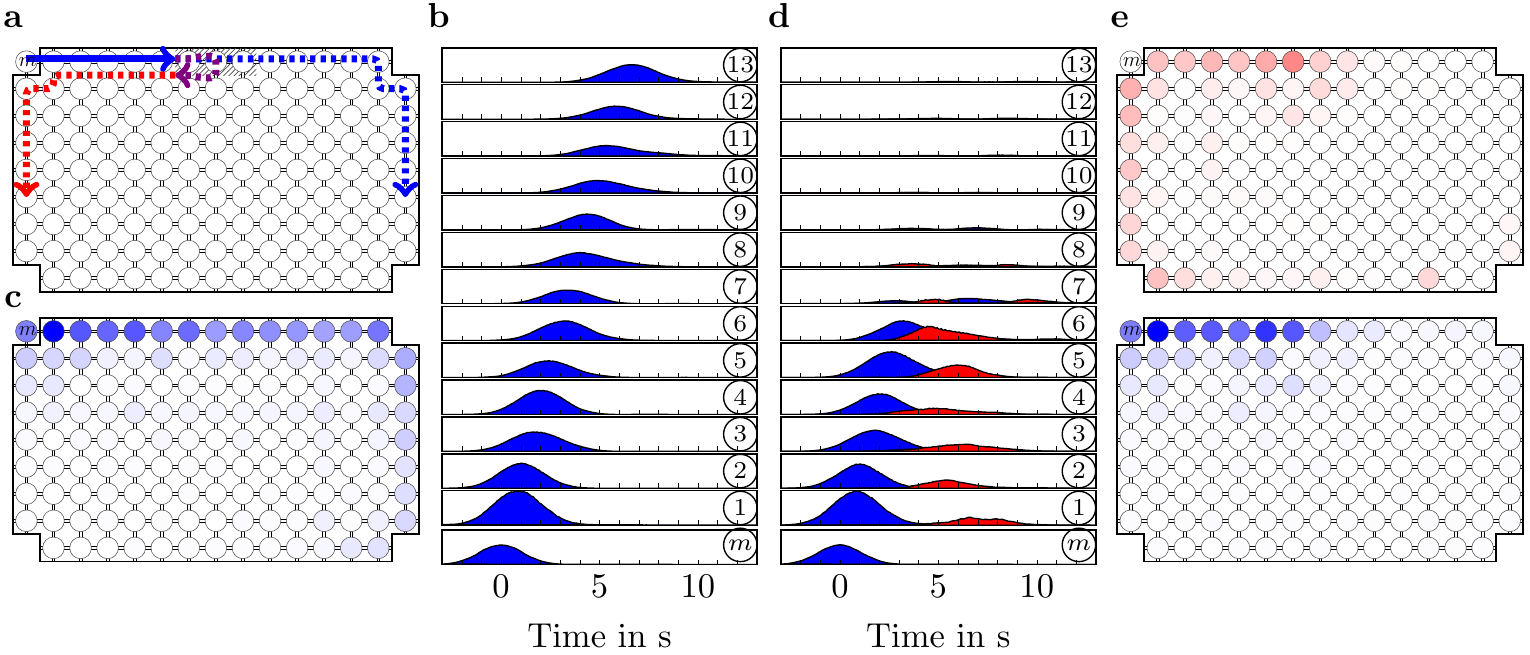}
	\caption{Propagation of a gaussian wave packet along the $r=0$ edge. The packet is induced on site $(0,0)$ and propagates along positive $s$ direction due to the chosen polarization. The results of the left (red) polarization is scaled by a factor of two with respect to the right (blue) polarization to enhance visibility. (a) Schematic of the expected behavior. Depending on the local edge potential, the wave packet passes through or is reflected. (b) and (d) Time, site and polarization resolved kinetic energy for the locally symmetry preserved (b) and symmetry broken (d) edge potentials. In (b) only the right polarization is shown. (c) and (e) Site and polarization resolved kinetic energy integrated over time. In (c) only the right polarization is shown. }
	\label{fig:wavepacket}
\end{figure*}

The drive can generate any motion on the driven site, in particular polarized excitations. Depending on the polarization, the edge channels transport energy only in one direction due to their chirality. Therefore, we drive the experiment once on site $(0,0)$ as shown in \fref{fig:setup} and once  on site $(8,0)$. This procedure allows to carry out the spectrum measurements polarization resolved. The results are shown in \fref{fig:measuredSpectra}. The three panels correspond to the calculated band structures shown in \fref{fig:spectra} (c-e).

The results show that we can recover bulk bands, bulk band gaps, edge bands and edge band gaps. The bulk band gaps were determined in the symmetry preserving configuration $\alpha=0.81$, while the edge band gaps where analyzed for each configuration separately. In addition, for the edge bands we extract the dispersion relations, sensitive to what polarization we used to excite the system. The experiments are restricted to the lower part of the spectrum because that is sufficient for what follows.
%
\section{Switching a phonon channel} 
\label{sec:switching_a_phonon_channel}
The previous two sections showed how to choose the edge potential to specifically shape the edge dispersions. In a next step we mean to combine two different potentials to engineer the desired switch. The basic idea is the following: Assume that we combine two semi-infinite systems, such that they have one single edge along the $s$ direction. For sites $s<0$, i.e., on the `negative' half space, we apply an edge potential of the form $\alpha^{x/y}=\alpha$, while on sites $s\geq 0$, the `positive' half space, we choose an edge potential $\alpha^x\neq \alpha^y$.

From the band structure calculations we know that for frequencies within the edge band gap of the positive half space there are no states which could transport energy through the system. If we have an incident wave coming from the negative half space towards the boundary, it has to be reflected at the interface. Furthermore, because the sign of the group velocity of the edge channels is bound to specific polarizations, the wave has to switch polarization while being reflected. 

Instead of considering two semi-infinite systems, we adapt the setup for practical purposes. We start with a system which has a symmetry preserved edge potential featuring no edge band gap. Within this system, to $n$ consecutive edge sites a symmetry-breaking potential is applied, opening an `edge band gap' locally. Due to the finite extend of the symmetry breaking potential, the reflection of incoming waves is not perfect anymore: evanescent waves can pierce through. However, the transmission is exponentially suppressed.

By selectively preserving or breaking the symmetry on this finite region, we can switch channels on and off. For the application as a switch it is desirable to have the region which needs to be modified as small as possible. In our experimental setup, we have chosen this region to be only about four sites long while we still keep an almost perfect suppression. Along the long edges ($r\in\{0,8\}$) we implement $\alpha=0.81$ on all sites. We then selectively change the couplings connecting sites $(0,5-9)$, shaded in \fref{fig:setup}, to be $\alpha^x=0.81$, $\alpha^y=0.24$. This locally breaks the protecting symmetry. 

To measure the performance of our switch we induce a polarized gaussian wave packet along the $r=0$ edge of the system. Its carrier frequency is chosen to be $\unit[2.22]{Hz}$, such that it lies well in the edge band gap of the symmetry broken configuration. First we leave the symmetry intact, then we break it as described above. In either case we track the time-resolved motion of all the sites. On every site we separate the motion into the two polarizations \footnote{The two polarizations always offer a basis to describe the motion, whether the symmetry is preserved or not.}. The time resolved kinetic energies per site and polarization along the $r=0$ edge are shown in \fref{fig:wavepacket} (b) and (d). Without the symmetry breaking part, the wave packet travels along the full edge. Once the symmetry is broken locally, the wave packet is scattered into the other polarization channel while being reflected.

Panels (b) and (d) show the time resolved motion of the $r=0$ edge, but none of the other sites. To show that nothing is missing on the other sites, we integrated the kinetic energy per site and polarization over the time span of the experiment and show the result in \fref{fig:wavepacket} (c) and (e). For the symmetry-preserved situation we only show one polarization as the other one is not excited. From these panels we see that the restriction to the edge sites is well justified.
%
\section{Conclusions} 
\label{sec:conclusions}
We have demonstrated the implementation of a switch for topological phonon channels. The channels were realized as topological edge modes of a two dimensional analogue of the quantum spin Hall effect. A crucial ingredient was that the underlying topology is protected by a local symmetry. By applying an edge potential, the edge dispersion can be altered, and in case that the potential breaks the protecting symmetry, a total band gap is opened. For frequencies within the total band gap, it is not possible to transport energy through the system.

At an interface between a symmetry-preserving and a symmetry-breaking potential, waves at frequencies in the total band gap will be reflected. This is the basic idea underlying the switch. The picture we engaged to explain our approach is based on band theory of infinite systems. However, we experimentally demonstrated that it is sufficient to apply the symmetry breaking potential on a very small region in space, being only of the order of one unit cell. Albeit our experiments where carried out in a mass spring model, the concept and strategy presented carry over to any physical implementation.
%
\appendix
\section*{Appendix}
\setcounter{section}{1}
Parameters chosen in the analysis are according to the nominal values realized in the experimental setup. The pendula making up the individual degrees of freedom have a moment of inertia $J\approx\unit[0.123]{kg\,m^2}$ and a bare eigenfrequency $\omega_{0}/2\pi\approx \unit[0.75]{Hz}$. The reference coupling strength of the springs is $M\approx\unit[5.10]{Nm}$, leading to an effective coupling strength, $f=M/J$, of $\sqrt{f}/2\pi\approx \unit[1.02]{Hz}$.

The coupling matrices are given by $\Omega =-[\omega_0^2+(1+\sqrt{3})f] \mathds{1}_{3\times 3}$ and
\begin{equation}\nonumber
	\begin{aligned}
		A
		&=\begin{pmatrix}
			-2 & 1 & 0 \\
			1 & -2 & 1 \\
			0 & 1 & -2
		\end{pmatrix},
		&B
		&=\begin{pmatrix}
			0 & 0 & 0 \\
			0 & 0 & 0 \\
			1 & 0 & 0
		\end{pmatrix}, \\
		C
		&=\frac{1}{2} \begin{pmatrix}
			2 & 0 & 0 \\
			0 & -1 & 0 \\
			0 & 0 & -1
		\end{pmatrix},
		&D
		&=\frac{\sqrt{3}}{2}  \begin{pmatrix}
			0 & 0 & 0 \\
			0 & 1 & 0 \\
			0 & 0 & -1
		\end{pmatrix}.
	\end{aligned}
\end{equation}
%
\bibliography{ref}
\end{document}